# Correlation energy of intrashell doubly excited states of heliumlike atoms

**V V Kavera**
Krasnodar, Russia

E-mail: vad1809@rambler.ru

**Abstract.** The analysis of correlation energy of the simplest first approximation of a variational method ( $VM_1$ ) for the intrashell states of two-electron atoms is the purpose of the present work. According to this work, the method $VM_1$ allows to divide energy of atom on Coulomb and additional not - Coulomb correlation parts with very high accuracy. Thus all the Coulomb interaction completely is taken into account in calculation and not - Coulomb terms enter only in correlation energy. The general formula for energy of additional correlation interaction is obtained and the constant of this interaction is determined. The algorithm of calculation of total energy of two-electron atoms is offered. The outcomes of calculations for intra-shell states are adduced and compared to the experimental data . The arguments for the capability of existence of the analytical solution for a three-body problem in atomic physics are presented. Some reasons about probable practical consequences are presented. Most interesting of them is the discovery that under certain conditions additional electron-electron attraction can exceed classical electron-electron repulsion.

## I. INTRODUCTION

The independent-particle model is widely used approach for the solution of a problem of many skew fields in physics generally and in atomic physics in particular. Just it is used at the description of x-ray and elementary optical spectra of atoms, at the description of a periodic system of elements and at the formulation of a principle Pauli.

Generally correlation energy of atom is determined as a residual between precision energy (taken from experiment) and energy calculated by one from methods using the independent-particle model. The Hartree-Fock method is used most frequently in this role, but term "correlation energy" can be applied also at use other methods.

The correlation energy of a Hartree-Fock method was a theme of many works at the end of 1950s and at the beginning of 1960s. However after appearance of the first experimental data on doubly excited states ( further - DES ) of atoms in a middle of 1960s the point of view has prevailed, that the Hartree-Fock method and generally the independent-particle model is inapplicable for the description of a similar class of states because of large increase of correlation energy for them ( see review [1] ). This conclusion was based mainly on the analysis known on that moment intershell $n_1 l_1 n_2 l_2$ DES of heliumlike atoms (where $n_1$ and $n_2$ - principal quantum numbers, and $l_1$ and $l_2$ - orbital quantum numbers of appropriate electrons).

For intrashell $nl_1 nl_2$ states (where $n$ – common principal quantum number for both electrons) the experimental data were known long time only for $n = 1$ (ground state 1s1s) and for $n = 2$. Only at the end of 1990s and beginning 2000s the sufficient data for $n = 3$ have appeared. Thus the capability to analyze relation of correlation energy to a principal quantum number $n$ and charge of nucleus $Z$ for intrashell states of heliumlike atoms has appeared only since that time.

The simplest first approximation of a variational method is used as the theoretical approach in the given work . The analysis of correlation energy of this approach for a ground state 1s1s of two-electron atoms can be found in various sources (for example in [2]), but it is not known to any publication of similar research for DES and it is the purpose of the present work.

## II. THEORETICAL APPROACH

Main problem at calculation of multielectron atoms, which elementary case are heliumlike atoms, eventually is the necessity to take into account of electron-electron interaction .

In the present work the one-configuration first approximation of a variational method (further VM$_1$) is used, in which one parameter - charge of the nucleus varies only, and all electron-electron interaction reduced only to mutual shielding of this charge. In difference from a Hartree-Fock method in a method VM$_1$ the application of a variational principle results in algebraic equations instead of differential equations, and the final results represent simple analytical expressions instead of tables. Thus VM$_1$ is most simple from all possible methods, which use independent-particle model and which allow to obtain a total separation of variables. Thus VM$_1$ is not one from many methods, but it is unique, since only it allows to take into account electron-electron interaction and at the same time completely to divide their variables, i.e. to receive the analytical solution of three-body problem.

The description VM$_1$ can be found in many sources (for example - in [2]) and there is no necessity to consider it here in details.

It is necessary to notice, that the exchange effects do not take into account in this approach. The wave function of two-electron atom in this approximation reads

$$\psi = \psi_1 \cdot \psi_2 , \qquad (1)$$

where $\psi_1$ and $\psi_2$ - wave functions of separate electrons, which differ from hydrogenlike functions only by replacement in them of a real charge of the nucleus $Z$ on effective charges $Z_{e1}$ and $Z_{e2}$, which are variational parameters of an variational equation

$$\iint \psi H \psi^* dV_1 dV_2 = \min , \qquad (2)$$

where $H$ – the nonrelativistic two-electron Hamiltonian reads in atomic units ( a.e.)

$$H = \frac{p_1^2 + p_2^2}{2} - \frac{Z}{r_1} - \frac{Z}{r_2} + \frac{1}{r_{12}} , \qquad (3)$$

where $r_1$ and $r_2$ are the electron-nucleus distances, $r_{12}$ is the electron-electron distance, whereas $p_1$ and $p_2$ are the individual momenta of the electrons.

Eventually formula of energy (hereinafter everywhere in the text Rydberg -Ry is used as units of energy) two-electron atom calculated by the method VM$_1$ looks like

$$E_{He1} = \varepsilon_1 \frac{Z_{e1}^2}{n_1^2} + \varepsilon_2 \frac{Z_{e2}^2}{n_2^2} , \qquad (4)$$

or for intrashell $nl_1nl_2$ states

$$E_{He1} = \frac{\varepsilon}{n^2}(Z_{e1}^2 + Z_{e2}^2) \qquad (5)$$

where $\varepsilon$ - factor, which take into account relativistic effects and influence of the nucleus. In the present work its values are taken from experiment with use of the formula

$$\varepsilon = E_{H\exp} \frac{n^2}{Z^2} , \qquad (6)$$

where $E_{H\exp}$ - experimental value of energy of hydrogenlike atom for given $Z$ and given $n$.

### III. CORRELATION ENERGY

The values $E_{He1}$, calculated by a method VM$_1$, values of energy $E_{He\,exp}$, taken from experiment, and also value of correlation energy $E_{cor}= E_{He\,exp} – E_{He1}$ are given in a table 1.

Table 1. The data of experiment and calculations for $nl_1 nl_2$ states of heliumlike atoms.

| State | Z | $E_{He\,exp}$ (Ry) | $E_{He1}$ (Ry) | $E_{cor}$ (Ry) | C | $E_{He}$ (Ry) | Ref. $E_{He\,exp}$ |
|---|---|---|---|---|---|---|---|
| *1s1s (1S)* | 1 | 1,0550 | 0,9448 | 0,1101 | 0,110 | 1,0559 | [3] |
| | 2 | 5,8068 | 5,6948 | 0,1120 | 0,112 | 5,8059 | [4] |
| | 3 | 14,5597 | 14,4457 | 0,1140 | 0,114 | 14,5568 | [4] |
| | 4 | 27,3131 | 27,1987 | 0,1143 | 0,114 | 27,3099 | [4] |
| | 5 | 44,0699 | 43,9561 | 0,1137 | 0,114 | 44,0672 | [4] |
| | 6 | 64,8318 | 64,7202 | 0,1116 | 0,112 | 64,8313 | [4] |
| | 7 | 89,6035 | 89,4946 | 0,1089 | 0,109 | 89,6057 | [4] |
| | 8 | 118,3845 | 118,2829 | 0,1016 | 0,102 | 118,3940 | [4] |
| | 9 | 151,1885 | 151,0900 | 0,0985 | 0,099 | 151,2011 | [4] |
| | 10 | 188,0111 | 187,9201 | 0,0910 | 0,091 | 188,0312 | [4] |
| *2s2s (1S)* | 1 | 0,2972 | 0,2444 | 0,0528 | 0,106 | 0,2999 | [5] |
| | 2 | 1,5571 | 1,4436 | 0,1135 | 0,114 | 1,5547 | [6] |
| | 3 | 3,8077 | 3,6431 | 0,1646 | 0,110 | 3,8098 | [7] |
| | 4 | 7,0788 | 6,8434 | 0,2354 | 0,118 | 7,0656 | [7] |
| | 5 | 11,3354 | 11,0452 | 0,2902 | 0,116 | 11,3229 | [8] |
| *3s3s (1S)* | 2 | 0,7206 | 0,6431 | 0,0776 | 0,116 | 0,7171 | [9] |
| | 6 | 7,4304 | 7,2260 | 0,2044 | 0,102 | 7,4482 | [10] |
| | 7 | 10,2379 | 9,9849 | 0,2530 | 0,108 | 10,2442 | [11] |
| | 8 | 13,4581 | 13,1900 | 0,2681 | 0,101 | 13,4863 | [10] |
| *2s2p (1P)* | 1 | 0,2520 | 0,2313 | 0,0207 | 0,083 | 0,2498 | [5] |
| | 2 | 1,3860 | 1,4071 | -0,0210 | 0,084 | 1,3700 | [12] |
| | 3 | 3,5121 | 3,5832 | -0,0711 | 0,095 | 3,4906 | [15] |
| | 4 | 6,6378 | 6,7600 | -0,1222 | 0,098 | 6,6119 | [7] |
| | 5 | 10,7842 | 10,9383 | -0,1542 | 0,088 | 10,7346 | [8] |
| *3s3p (1P)* | 1 | 0,1249 | 0,1062 | 0,0187 | 0,112 | 0,1247 | [5] |
| | 2 | 0,6712 | 0,6353 | 0,0359 | 0,108 | 0,6723 | [12] |
| *4s4p (1P)* | 2 | 0,3888 | 0,3595 | 0,0294 | 0,117 | 0,3872 | [12] |
| *5s5p (1P)* | 1 | 0,0491 | 0,0390 | 0,0101 | 0,101 | 0,0501 | [5] |
| | 2 | 0,2528 | 0,2307 | 0,0221 | 0,110 | 0,2529 | [12] |
| *6s6p (1P)* | 1 | 0,0348 | 0,0272 | 0,0076 | 0,091 | 0,0364 | [5] |
| | 2 | 0,1760 | 0,1605 | 0,0156 | 0,094 | 0,1790 | [12] |
| *2s2p (3P)* | 1 | 0,2841 | 0,2313 | 0,0528 | 0,106 | 0,2869 | [5] |
| | 2 | 1,5218 | 1,4071 | 0,1147 | 0,115 | 1,5182 | [6] |
| | 3 | 3,7536 | 3,5832 | 0,1704 | 0,114 | 3,7498 | [7] |
| | 4 | 6,9906 | 6,7600 | 0,2306 | 0,115 | 6,9823 | [7] |
| | 5 | 11,2292 | 10,9383 | 0,2909 | 0,116 | 11,2161 | [13] |
| | 6 | 16,4630 | 16,1188 | 0,3442 | 0,115 | 16,4521 | [13] |
| | 10 | 47,4169 | 46,8875 | 0,5294 | 0,106 | 47,4431 | [13] |
| | 12 | 68,9347 | 68,3223 | 0,6125 | 0,102 | 68,9889 | [13] |
| *3s3p (3P)* | 2 | 0,6988 | 0,6353 | 0,0635 | 0,095 | 0,7094 | [14] |
| | 7 | 10,2085 | 9,9533 | 0,2552 | 0,109 | 10,2126 | [11] |
| *2p2p (1S)* | 2 | 1,2454 | 1,3393 | -0,0939 | 0,094 | 1,1171 | [6] |
| | 3 | 3,2578 | 3,4764 | -0,2186 | 0,109 | 3,1430 | [7] |
| | 4 | 6,2414 | 6,6141 | -0,3727 | 0,124 | 6,1697 | [7] |
| | 5 | 10,2991 | 10,7533 | -0,4542 | 0,114 | 10,1978 | [8] |
| *3p3p (1S)* | 2 | 0,6354 | 0,6246 | 0,0108 | - | 0,6246 | [9] |
| *2p2p (1D)* | 1 | 0,2561 | 0,2026 | 0,0535 | 0,107 | 0,2582 | [5] |
| | 2 | 1,4049 | 1,3393 | 0,0656 | 0,131 | 1,3949 | [6] |
| | 3 | 3,5376 | 3,4764 | 0,0612 | 0,122 | 3,5319 | [7] |
| | 4 | 6,6744 | 6,6141 | 0,0603 | 0,121 | 6,6697 | [7] |
| | 5 | 10,8149 | 10,7533 | 0,0616 | 0,123 | 10,8089 | [13] |
| | 6 | 15,9391 | 15,8947 | 0,0444 | 0,089 | 15,9502 | [13] |
| *3p3p (1D)* | 2 | 0,6949 | 0,6246 | 0,0703 | 0,105 | 0,6987 | [9] |
| | 6 | 7,3503 | 7,1638 | 0,1865 | 0,093 | 7,3860 | [10] |
| | 7 | 10,1497 | 9,9118 | 0,2379 | 0,102 | 10,1711 | [11] |
| | 8 | 13,3589 | 13,1059 | 0,2529 | 0,095 | 13,4022 | [10] |
| *2p2p (3P)* | 2 | 1,4205 | 1,3393 | 0,0812 | 0,122 | 1,4134 | [13] |
| | 4 | 6,7668 | 6,6141 | 0,1527 | 0,115 | 6,7623 | [13] |
| | 5 | 10,9450 | 10,7533 | 0,1917 | 0,115 | 10,9385 | [13] |
| | 6 | 16,1096 | 15,8947 | 0,2149 | 0,107 | 16,1169 | [13] |
| | 8 | 29,5400 | 29,1887 | 0,3514 | 0,132 | 29,4849 | [13] |
| *3p3p (3P)* | 7 | 10,0615 | 9,9118 | 0,1497 | 0,128 | 10,0415 | [11] |
| *3d3d (1D)* | 2 | 0,6545 | 0,5780 | 0,0765 | 0,115 | 0,6520 | [9] |
| | 6 | 7,2040 | 7,0037 | 0,2004 | 0,100 | 7,2259 | [10] |
| | 7 | 10,0101 | 9,7233 | 0,2868 | 0,123 | 9,9826 | [11] |
| | 8 | 13,2090 | 12,8890 | 0,3200 | 0,120 | 13,1853 | [10] |
| *3d3d (1G)* | 2 | 0,6133 | 0,5779 | 0,0353 | 0,106 | 0,6150 | [14] |

The analysis of relation of correlation energy from $Z$ and $n$ shows that the $nl_1 nl_2$ states are divided into two groups. The first group involves lowest states for the given configuration, i.e. states with lowest possible $n$. It is states, in which at least one electron is on the orbit, for which $n = l + 1$. In the old quantum theory this condition corresponds to the special case of circular orbits. Second group involves all remaining states of configurations, i.e. states with $n > l + 1$ for everyone from two electrons. The formulas were obtained

$$E_{cor} = \frac{C}{2} \cdot \frac{Z}{n}[k_1 - k_2 \frac{(Z-1)}{Z}] \qquad (7)$$

and

$$E_{cor} = \frac{C}{2} \cdot \frac{Z}{n} k_1 \qquad (8)$$

accordingly for the first and second group of states, where $k_1$ and $k_2$ - integer factors, which values are given in a table 2, and $C$ - numerical factor. The values $C$, at which formulas (7) and (8) give precision coincidence with experiment, are given in a table 1. It is easy to see, taking into account errors of measurements, that $C$ is a constant, which average value is close to 1/9.

Table 2. Factors $k_1$ and $k_2$.

| State | $k_1$ | $k_2$ |
|---|---|---|
| *nsns(1S)* | 2 | 2 |
| *npnp(1D)* | 2 | 2 |
| *nsnp(3P)* | 2 | 0 |
| *npnp(3P)* | 1 | 0 |
| *nsnp(1P)* | 1 | 3 |
| *npnp(1S)* | 0 | 4 |

In that specific case of states *nsns (1S), npnp (1D), ndnd (1G)* etc., when both electrons are on same orbit, or, in other words, occupy the same quantum cell, the formulas (7) and (8) become especially simple

$$E_{cor} = C \frac{1}{n} \qquad (9)$$

for $n = l + 1$ and

$$E_{cor} = C \frac{Z}{n} \qquad (10)$$

for $n > l + 1$.

It is necessary to notice, that refusal to consider of exchange degeneration is justified for these states from any point of view.

The particular case of the formula (9) for $n = 1$ corresponds to a ground state of heliumlike atoms (1s1s) and results to $E_{cor} = C$, i.e. the correlation energy in this case does not depend from $Z$ and it was noted by Bethe as the curious fact in [2]. Except for a Bethe this fact (and it is only particular case from (7) - (10)) attracted some more other investigators in 1950s and 1960s. Actually main difference of the given article from those activities is an application same well known, but a little forgotten, method of calculation $VM_1$ to the newest experimental data on doubly excited states. If these data were known in 1950s-1960s, the method $VM_1$ would not be forgot, and the formulas (7) - (10), would be obtained still then.

The formulas obtained above shows, that at least in case of a method $VM_1$ the independent-particle model can be applied successful for doubly excited states if to consider correlation energy not as an annoying error, but as the simply taken into account correction with interesting physical properties which will be discussed below. Moreover, simplicity of obtained evaluations results that the correlation energy is transformed from a problem into a proof of effectiveness of the independent-particle model.

Already now it is possible to use formulas (7-10) to describe known lines of spectra of heliumlike atoms and to predict or to help to identify unknowns lines.

The semi-empirical formula was finally obtained for calculation of total energy of heliumlike atom for $nl_1nl_2$ states

$$E_{He} = E_{He1} + E_{cor} , \qquad (11)$$

where $E_{He1}$ is calculated by a method $VM_1$ and follows from the formula (5), and $E_{cor}$ follows from the formulas (7) and (8) and is entered because of analysis of the experimental data. The data of calculation of energy $E_{He}$ under the formula (11), in the supposition, that $C = 1/9$, are given in table 1 for those of $nl_1nl_2$ states, for which the experimental data are known . Accordance of calculations and experiment quite satisfactory, considering errors of measurements and approximations at calculations. The data of the same calculation for a unknown for today $nsns$ (1S) states at $n$ from 4 up to 10 and $Z$ from 1 up to 10 are given as an example in a table 3.

**Table 3. The data of calculation for *nsns* (1S) states for *n* from 4 up to 10.**

| Z  | $E_{He}$(Ry) | $E_{He}$(Ry) | $E_{He}$(Ry) | $E_{He}$(Ry) | $E_{He}$(Ry) | $E_{He}$(Ry) | $E_{He}$(Ry) |
|----|--------------|--------------|--------------|--------------|--------------|--------------|--------------|
|    | *4s4s*       | *5s5s*       | *6s6s*       | *7s7s*       | *8s8s*       | *9s9s*       | *10s10s*     |
| 1  | 0,0893       | 0,0616       | 0,0459       | 0,0360       | 0,0293       | 0,0245       | 0,0210       |
| 2  | 0,4175       | 0,2762       | 0,1980       | 0,1500       | 0,1183       | 0,0962       | 0,0802       |
| 3  | 0,9958       | 0,6508       | 0,4612       | 0,3457       | 0,2699       | 0,2174       | 0,1794       |
| 4  | 1,8242       | 1,1854       | 0,8356       | 0,6230       | 0,4839       | 0,3879       | 0,3186       |
| 5  | 2,9029       | 1,8802       | 1,3212       | 0,9820       | 0,7605       | 0,6078       | 0,4979       |
| 6  | 4,2319       | 2,7352       | 1,9180       | 1,4227       | 1,0997       | 0,8771       | 0,7170       |
| 7  | 5,8114       | 3,7504       | 2,6260       | 1,9452       | 1,5014       | 1,1959       | 0,9763       |
| 8  | 7,6417       | 4,9261       | 3,4455       | 2,5494       | 1,9657       | 1,5641       | 1,2758       |
| 9  | 9,7230       | 6,2623       | 4,3763       | 3,2355       | 2,4927       | 1,9818       | 1,6152       |
| 10 | 12,0556      | 7,7592       | 5,4183       | 4,0035       | 3,0823       | 2,4490       | 1,9943       |

### IV. RESULTS AND COMMENTS

4.1. From a point of view of physics the formulas (7-10) result to unusual, i.e. in non-classical relation of energy of interaction of charged particles to a distance between them.

If to present $n$ as a radius of atom $r$ (remembering, that in hydrogenlike atoms $r \sim n^2$), it is easy to obtain for different parts of full energy of atom of relation

$E \sim 1/r^k$ ,

where for terms, calculated by the method $VM_1$, $k = 1$, that completely corresponds to the classical law of the Coulomb, and for correlation energy $k = 1/2$. Thus here there is an additional interaction decreasing on a distance slower, than Coulomb force.

The additional calculations have shown, that the relation $E \sim 1/r^{1/2}$, or accordingly $E \sim 1/n$, occurs only in that case, when both electrons have identical principal quantum numbers $n$. The author have proofs of existence of similar relation of correlation energy from a principal quantum number of external electrons not only in heliumlike atoms, but also in atoms with large number of electrons, and in molecules and crystals.

That fact, that coincidence of principal quantum numbers $n$ at both electrons is required for appearance of relation $E \sim 1/n$, can indicate resonant character of additional interaction. Moreover it results in electron-electron attraction, instead of repulsion and very strongly depends on a configuration of spin and orbital moment, that makes it even less similar to electrostatic interaction, but similar to interaction, which exist between protons in the nucleus.

4.2. From a mathematical point of view the method $VM_1$ allows to divide energy of atom on Coulomb and not - Coulomb parts with very high accuracy. Thus all the Coulomb interaction completely is taken into account in calculation and not - Coulomb terms enter only in correlation energy. It is correctly not only for intrashell states, but also generally for all types of states of heliumlike atoms. Thus, if the not - Coulomb interactions were absent in two-electron atom, the method $VM_1$ would give the precisiouly analytical solution of a three-body problem in atomic physics.

The simplicity of the formulas (7-10) allows to hope, that the analytical solution is possible and with considering of not - Coulomb interactions. It would become possible after an evaluation of correlation energy $E_{cor}$ and constant $C$ from certain general principles.

4.3. From a practical point of view it is interesting, that since some value $n$ usual Coulomb repulsion electrons (decreasing as $1/n^2$) will become less additional not - Coulomb attractions (decreasing as $1/n$). It can result in macroscopic case to join of electrons in certain stable or metastable structures just as the protons are integrated in the nucleus. The similar processes could spontaneously happen in nature. It is possible, that the similar effects could explain at least some from anomalous plasma-like effects observed in atmosphere and an ionosphere.

To receive a similar new states of substance in experiment, it is necessary, that the electrons of substance were excited synchronously, i.e. had identical energy and identical values $n$ in each instant . To the present moment not much of similar (doubly exited) states is obtained even for two-electron atoms. It is known them even less for molecules. Moreover both in case of atoms, and in case of molecules the values $n$ are not reached yet value, at which the electron-electron attraction exceeds a repulsion. In case of macroscopic skew fields the problem of synchronous excitation of electrons up to maximum large $n$ till now not to pose, though technically it is possible.

Let's remind also, that the explanation of a superconductivity involves appearance of additional electron-electron attraction , which exceeds Coulomb repulsion under certain conditions. Moreover there are the direct analogies between additional correlation energy of electrons in superconductors and additional correlation energy of electrons in separate atoms and the Cooper pair sometimes is represented as two electrons moving round an induced positive charge, and is compared to atom of a helium. All of this makes probable bose-einstein condensation of synchronously excited electrons both in atoms, and in macroscopic skew fields from that moment, when not - Coulomb electron-electron attraction will begin to exceed Coulomb repulsion. The similar superconductivity already could be named super-high-temperature.

## V. CONCLUSION

The approach based on separation of total energy of two-electron atoms on classical Coulomb and nonclassical correlation parts, allows on the one hand to simplify calculations, and with another - to see interesting regularities, which were not visible at use of more complex methods. Most interesting is the discovery of that fact, that under certain conditions electron-electron attraction exceeds electron-electron repulsion. The most important practical consequence it is the capability of existence of ordered structures of a new type in the special way exited substance.